\documentclass[apj,square,numbers]{emulateapj}
\usepackage{graphicx}
\usepackage{xcolor}
\usepackage{amsmath,amssymb,latexsym}
\usepackage{hyperref}
\usepackage{ulem}
\begin{document}
\title{Revisiting Studies of the Statistical Property of a Strong Gravitational Lens System and Model-independent Constraint on the Curvature of the Universe}

\author{Jun-Qing Xia$^{1,\ast}$, Hai Yu$^{2,3,1}$, Guo-Jian Wang$^{1}$, Shu-Xun Tian$^{1}$, Zheng-Xiang Li$^{1}$, Shuo Cao$^{1}$, Zong-Hong Zhu$^{1}$}

\affil{$^{1}$Department of Astronomy, Beijing Normal University, Beijing 100875, China; xiajq@bnu.edu.cn}
\affil{$^{2}$School of Astronomy and Space Science, Nanjing University, Nanjing 210093, China}
\affil{$^{3}$Key Laboratory of Modern Astronomy and Astrophysics (Nanjing University), Nanjing 210093, China}

\begin{abstract}

In this paper we use a recently compiled data set, which comprises 118 galactic-scale strong gravitational lensing (SGL) systems to constrain the statistic property of SGL system, as well as the curvature of universe without assuming any fiducial cosmological model. Based on the singular isothermal ellipsoid (SIE) model of SGL system, we obtain that the constrained curvature parameter $\Omega_{\rm k}$ is close to zero from the SGL data, which is consistent with the latest result of planck measurement. More interestingly, we find that the parameter $f$ in the SIE model is strongly correlated with the curvature $\Omega_{\rm k}$. Neglecting this correlation in the analysis will significantly overestimate the constraining power of SGL data on the curvature. Furthermore, the obtained constraint on $f$ is different from previous results: $f=1.105\pm0.030$ ($68\%$ C.L.), which means that the standard singular isothermal sphere (SIS) model ($f=1$) is disfavored by the current SGL data at more than $3\sigma$ confidence level. We also divide the whole SGL data into two parts according to the centric stellar velocity dispersion $\sigma_{\rm c}$ and find that the larger value of $\sigma_{\rm c}$ the subsample has, the more favored the standard SIS model is. Finally, we extend the SIE model by assuming the power-law density profiles for the total mass density, $\rho=\rho_0(r/r_0)^{-\alpha}$, and luminosity density, $\nu=\nu_0(r/r_0)^{-\delta}$, and obtain the constraints on the power-law indexes: $\alpha=1.95\pm0.04$ and $\delta=2.40\pm0.13$ at 68\% confidence level. When assuming the power-law index $\alpha=\delta=\gamma$, this scenario is totally disfavored by the current SGL data, $\chi^2_{\rm min,\gamma} - \chi^2_{\rm min,SIE} \simeq 53$.

\end{abstract}


\maketitle

\section{Introduction}\label{sec:intro}

The curvature of universe, which is often parameterized by $\Omega_{\rm k}$, is a fundamental parameter in cosmology. It determines whether our universe is open ($\Omega_{\rm k}>0$), flat ($\Omega_{\rm k}=0$) or close ($\Omega_{\rm k}<0$). Currently, most of cosmological observations favor that $\Omega_{\rm k}$ is very
close to zero, such as the latest constraint from the Planck measurement, $|\Omega_K|<0.005$ \citep{Planck2015}. However, these constrains on $\Omega_{\rm k}$ are obtained by using a
model-dependent method. Therefore, constraints on curvature using the model-independent method is still very attractive in the literature \citep{Bernstein2006,Clarkson2007,Oguri2012,Li2014,Rasanen2015,Cai2016,Yu2016}. For instance, in \citet{Bernstein2006}, based on the basic distance sum rule, they used the weak lensing to test the curvature in a model-independent way. Recently, this method was also used in \citet{Rasanen2015} with the strong gravitational lensing (SGL) systems to test the curvature of universe and the Friedmann-Lema\^{\i}tre-Robertson-Walker (FLRW) metric. \citet{Rasanen2015} collected 38 SGL systems and used them to constrain $\Omega_{\rm k}$ and found that the SGL sample gives the consistent constraint with the flat universe although the statistical error is very large, $-1.22<\Omega_{\rm k}<0.63$ at 95\% confidence level.

As an important prediction of the General Relativity (GR), SGL has become a powerful tool to test cosmology, astrophysics (the structure, formation, and evolution of galaxies and galaxy clusters), and the gravity theories. The first discovery of the SGL system Q0957+561 \citep{Walsh1979} hints us the possibility of using
the galactic lensing systems in the study of cosmological parameters and the galaxy structure. In a specific strong-lensing system, the background source (high-redshift quasar, supernova or galaxy) will reveal itself as multiple images, due to the gravitational field of the intervening lens (usually an elliptical galaxy) between the observer and the source. Meanwhile, as the image separation in the system depends on angular diameter distances to the lens and to the source, the observation of SGL can provide the information of a distance ratio ${d_{\rm ls}}/{d_{\rm s}}$, where $d_{\rm ls}$ is the angular diameter distance between the source and lens and $d_{\rm s}$ is that between the the source and observer. Following this direction, many recent works provided successful applications of different SGL samples in the investigation of the structure and evolution of galaxies \citep{Zhu1997,Treu2006,Cao2016}, the post-Newtonian parameter describing the deviations from the GR \citep{Schwab10}, the dynamical properties of dark energy \citep{Zhu2000,Chae2004,Cao2015,Li2016}, and the curvature of our universe \citep{Bernstein2006,Rasanen2015}.

Furthermore, precise spectroscopic and astrometric observations, obtained with well-defined selection criteria, may help us to study the statistic properties of lensing galaxies. Compared with late-type and unknown-type galaxies, early-type galaxies (or elliptical galaxies) which contains most of the stellar mass in the universe, provide most of the lensing galaxies in the available galactic SGL systems. Therefore, the singular isothermal ellipsoid (SIE) model, which has a elliptical projected mass distribution \citep{Ratnatunga1999}, and the singular isothermal sphere (SIS) model, which has a spherical symmetrical mass distribution, are two useful assumptions and good first-order approximations in statistical gravitational lensing studies. For a SIE lens, the Einstein radius of a SGL system can be calculated by the theoretical expression that
\begin{equation}\label{EinsteinRadius}
\theta_{\rm E} = 4\pi\frac{f^2\sigma_{\rm c}^2}{c^2}\frac{d_{\rm ls}}{d_{\rm s}}~,
\end{equation}
where $\theta_{\rm E}$ is the Einstein radius, $\sigma_{\rm c}$ is the central velocity dispersion of the lensing galaxy , $c$ is the speed of light and $f$ is a phenomenological coefficient which includes several systematic errors caused by: the difference between the observed stellar velocity dispersion and that of the SIS model, the assumption of SIS model in calculating the theoretical Einstein radius, and the softened isothermal sphere potentials \citep{Ofek2003,Cao2012}. For the standard SIS model, the coefficient parameter reduces to $f=1$. In order to take the uncertainty of $f$ into account, a prevalent procedure in the literature is to directly include a prior on $f^2$ with the $20\%$ uncertainty in the analysis. This method was firstly introduced by \citet{Kochanek2000,Ofek2003}, and extensively applied in the recent works by \citet{Rasanen2015,Holanda2016}. However, the statistical feature of the parameter $f$ is still not well understood.

Recently, based on a gravitational lens dataset including 70 galactic systems from Sloan Lens ACS (SLACS) and Lens Structure and Dynamics survey (LSD), \cite{Cao2012} treated $f$ as a free-parameter to fit the matter energy density $\Omega_{\rm m}$ and the equation of state of dark energy $w$. In the framework of a flat universe, their results showed that on the statistical level the 68\% C.L. constraint is $f^2=1.01\pm0.02$. However, in a more recent work by \citet{Rasanen2015}, the authors found that the parameter $f$ might be correlated with the cosmic curvature $\Omega_{\rm k}$, although they did not fully take the uncertainty of $f$ into account properly. Given the availability of a new sample of 118 lenses \citep{Cao2015} observed by the Sloan Lens ACS Survey (SLACS), BOSS emission-line lens survey (BELLS), Lens Structure and Dynamics (LSD), and Strong Lensing Legacy Survey (SL2S), the purpose of this work is to reconsider the studies on the curvature from this latest SGL system and fully consider the effect of the uncertainty of the parameter $f$ in the analysis. The structure of this paper is organized as follows. In Sec.~\ref{sec:method} we will introduce the
method and the SGL data used in this work. Then we will show our numerical result in Sec.~\ref{sec:result}. Finally, some discussion and summary will be in Sec.~\ref{sec:discussion}.

\section{Method and Data}\label{sec:method}

\subsection{Distance Sum Rule Method}

The basic distance sum rule is a simple geometric rule. Imaging that there are three points $A$, $B$ and $C$ on a straight line and $B$ lies between $A$ and $C$, the distances among them are $d_{\rm AB}$, $d_{\rm AC}$ and $d_{\rm BC}$, respectively. Then, obviously, there is a relation that $d_{\rm AC}=d_{\rm AB}+d_{\rm BC}$. However, if this line is not straight, the relation will become invalid. For example, assuming the line is a part of arc, then we have $d_{\rm AC}<d_{\rm AB}+d_{\rm BC}$. This rule is the same in the universe. Considering a SGL system in the universe, the distances between the observer and the lens galaxy and the sources are $d_{\rm l}$ and $d_{\rm s}$, respectively, while the distance between the lens galaxy and the source is $d_{\rm ls}$. Therefore, we have the equation $d_{\rm s}=d_{\rm l}+d_{\rm ls}$ if our universe is flat ($\Omega_{\rm k}=0$). Otherwise, there is $d_{\rm s}>d_{\rm l}+d_{\rm ls}$ or $d_{\rm s}<d_{\rm l}+d_{\rm ls}$ for $\Omega_{\rm k}<0$ or $\Omega_{\rm k}>0$, respectively (see figure 1 of \citet{Bernstein2006} for an illustration).

In the FLRW metric, the dimensionless comoving angular diameter distance between the lens galaxy, $z_{\rm l}$, and the source, $z_{\rm s}$, on a certain direction can be expressed as
\begin{eqnarray}\label{dcadd}
d(z_{\rm l},z_{\rm s}) &=& (1+z_{\rm s})H_0D_{\rm A}(z_{\rm l},z_{\rm s}) \nonumber\\
&=& \frac{1}{\sqrt{|\Omega_{\rm k}|}}{\rm sinn}\left[\sqrt{|\Omega_{\rm k}|}\int_{z_{\rm s}}^{z_{\rm l}}\frac{dz'}{E(z')}\right]~,
\end{eqnarray}
where ${\rm sinn}(x\sqrt{ |\Omega_{\rm k}| })/\sqrt{ |\Omega_{\rm k}| }=\sin(x),\,x,\,\sinh(x)$ if $\Omega_{\rm k}<0,\,=0,\,>0$, respectively. $E(z)=H(z)/H_0$ is the reduced Hubble parameter at redshift $z$ and $H_0$ is the Hubble constant. Therefore, $d_{\rm l}$, $d_{\rm s}$ and $d_{\rm ls}$ are equal to $d(z_{\rm l})=d(0,z_{\rm l})$, $d(z_{\rm s})=d(0,z_{\rm s})$ and $d(z_{\rm l},z_{\rm s})$, respectively. Based on Equation (\ref{dcadd}), one can derive the ratio of $d_{\rm ls}$ and $d_{\rm s}$ as the function of $\Omega_{\rm k}$, $d_{\rm l}$ and $d_{\rm s}$ \citep{Peebles1993,Rasanen2015}:
\begin{equation}\label{distanceRatio_th}
\frac{d_{\rm ls}}{d_{\rm s}} = \sqrt{1+\Omega_{\rm k} d_{\rm l}^2}-\frac{d_{\rm l}}{d_{\rm s}}\sqrt{1+\Omega_{\rm k} d_{\rm s}^2}.
\end{equation}
Therefore, if we obtain the direct distance information of $d_{\rm l}$, $d_{\rm s}$ and $d_{\rm ls}$ from the observations, the determination of the curvature $\Omega_{\rm k}$ can be derived straightforwardly from the observational data using this equation without introducing any fiducial model.

Fortunately, we can extract the distance information of ${d_{\rm ls}}/{d_{\rm s}}$ directly from the observation of the SGL system through the Equation (\ref{EinsteinRadius}):
\begin{equation}\label{distanceRatio_obs}
\frac{d_{\rm ls}}{d_{\rm s}} = \frac{\theta_{\rm E} c^2}{4\pi f^2\sigma_{\rm c}^2}~.
\end{equation}
In this paper, we used a comprehensive compilation of 118 strong lensing systems observed by four surveys: SLACS, BELLS, LSD and SL2S, which is also the largest gravitational lens sample published in the recent work \citep{Cao2015}. The SLACS data comprise 57 strong lenses presented in \citet{Bolton2008,Auger2009}, the BELLS
data comprise 25 lenses taken from \citet{Brownstein2012}, then 5 most reliable lenses from the LSD survey were taken from \citet{Koopmans2003,Treu2002,Treu2004}, and the SL2S data for a total of 31 lenses were taken from \citet{Sonnenfeld2013a,Sonnenfeld2013b}.

The Einstein radius $\theta_E$ is defined to be the radius at which the mean surface mass density $\Sigma$ equals the critical density $\Sigma_{cr}$ of the lensing configuration. However, for the lens galaxies one needs to assume an specific lens model to obtain the measurements of Einstein radii from observed strong lens systems in various lensing surveys. For all of the lenses from LSD, SLACS, BELLS and SL2S used in this paper, the resulted Einstein radii were obtained on the base of a singular isothermal ellipsoid (SIE) lens-mass model. Compared with the singular isothermal sphere (SIS) counterpart, SIE includes a two-dimensional potential of similar concentric and aligned elliptical iso-density contours, with minor-to-major axis ratio $q_{\rm SIE}$ \citep{Kassiola93,Kormann94,Keeton98}. We remark here that, although the resulted Einstein radii are slightly model dependent and SIE might not be accurate enough for some cosmological applications \citep{Saha00,Rusin02}, the assumption of an SIE model does not significantly bias the determination of the value of Einstein radii used in our analysis (See \citet{Sonnenfeld2013a,Sonnenfeld2013b} for details).

In this SGL data, they provided 118 SGL systems with detailed information about the redshift of lens and source galaxies $z_{\rm l}$ and $z_{\rm s}$, the Einstein Radius $\theta_{\rm E}$ and the central stellar velocity dispersion $\sigma_{\rm c}$. The lens galaxies of these 118 SGL systems spread in redshift range $0.075\leq z \leq1.004$ and the source galaxies spread in redshift range $0.196\leq z \leq3.595$. In practice, we subtract those SGL systems with $z_{\rm s}>1.4$, because the maximal redshift of the supernovae dataset we are using to determinate the cosmological distance is about $1.4$ (see details in the next subsection). Finally, there are 83 SGL systems (20 samples from BELLS, 57 samples from SLACS and 6 samples from SL2S) left in our sample.

\subsection{Determination the Distances}

In order to constrain the curvature of universe using the Equation (\ref{distanceRatio_th}), we still need to have the distance information of $d_{\rm l}$, $d_{\rm s}$, besides the distance ratio ${d_{\rm ls}}/{d_{\rm s}}$. To avoid the model-dependence, here we do not use the standard $\Lambda$CDM model to calculate the dimensionless comoving angular diameter distance information. Instead, we use the current observation of the type Ia supernova (SN) to determine the distance of lens galaxy $d_{\rm l}$ and the source galaxy $d_{\rm s}$. \citet{Suzuki2012} provided the SNIa Union2.1 compilation of 580 dataset from the Hubble Space Telescope Supernova Cosmology Project. The data are usually presented as tabulated distance modulus with errors. In this catalog, the redshift spans $0 < z < 1.414$, and about 95\% samples are in the low redshift region $z < 1$. The authors also provided the covariance matrix of data with and without systematic errors. In order to be conservative, we include systematic errors in our calculations.

Each SN sample in the Union 2.1 compilation gives the redshift $z$ and the luminosity distance $d_{\rm L}$. Using the relation between comoving angular diameter distance and luminosity distance, we can obtain that
\begin{equation}
d(z) = \frac{H_0d_{\rm L}(z)}{c(1+z)}~.
\end{equation}
It should be pointed out that the parameters of SN sample in Union 2.1 compilation are fitted with the parameters of cosmological model simultaneously. Therefore, the distance information of SN data are dependent on the input cosmological model. However, this effect caused by the assumed cosmological model can be omitted compared with the current larger uncertainties in modeling the SGL systems \citep{Rasanen2015}. Therefore, we simply use the distances and the uncertainties of the Union2.1 compilation reported in \citet{Suzuki2012}.

There are several methods to use the SN data to calibrate the distance, such as the Pad\'{e} approximation of order $(3,2)$ \citep{Liu2015,Lin2016} and the linear or cubic interpolation method \citep{Liang2008,Wang2016}. In this work, we use a simple third-order polynomial function with constraint conditions that $d(0)=0$ and $d^\prime(0)=1$ to fit the distance information of SN. It can be expressed as
\begin{equation}
d(z)=z+a_1z^2+a_2z^3~,
\end{equation}
where $a_{\rm i}$ are two free parameters which can be constrained from the SN Union2.1 compilation. Consequently, we can get the distance information of $d_{\rm l}$ and $d_{\rm s}$ from the SN observational data. Going from the third-order case to the fourth-order case does not improve the goodness of fit, which indicates that the third-order polynomial function is sufficient.

\section{Result}\label{sec:result}

In our analysis, we perform a global fitting using the COSMOMC package \citep{cosmomc}, a Monte Carlo Markov chain (MCMC) code. In the program, we make several modifications which allow us using the SGL data to constrain the curvature of universe. In the calculations, we have some free parameters which should be constrained from the SGL and SN datasets simultaneously:
\begin{equation}
\mathbf{P}=(a_1,a_2,\Omega_{\rm k},{\rm SGL}_{\rm i})~,
\end{equation}
where ${\rm SGL}_{\rm i}$ denote the free parameter of the specific model of the SGL system. Furthermore, in the FLRW framework, if the curvature $\Omega_{\rm k}<0$ the space will be a hypersphere and the comoving angular diameter distance will have an upper limit $d(z)\leq1/\sqrt{-\Omega_{\rm k}}$. The current observation of cosmic microwave background radiation (CMBR) gives the angular diameter distance at $z=1090$ is about $D_{\rm A}(1090)=12.8\pm0.07\,$Mpc \citep{Vonlanthen2010,Audren2013,Audren2014}. In the meanwhile we also have the direct probe on the current Hubble constant $H_0$ obtained from the re-analysis of \citet{riess2011} Cepheid data made by \citet{Efstathiou2014} by using a revised geometric maser distance to NGC 4258 from \citet{reid2013}: $H_0=72.5\pm2.5\,{\rm km/s/Mpc}$. Therefore, we add a prior about the lower limit of the curvature:
\begin{equation}
\Omega_{\rm k} \geq -\frac{c}{D_{\rm A}(1090)(1+1090)H_0} \simeq -0.1~.
\end{equation}

\begin{figure}[t]
    \centering
    \includegraphics[width=0.5\textwidth]{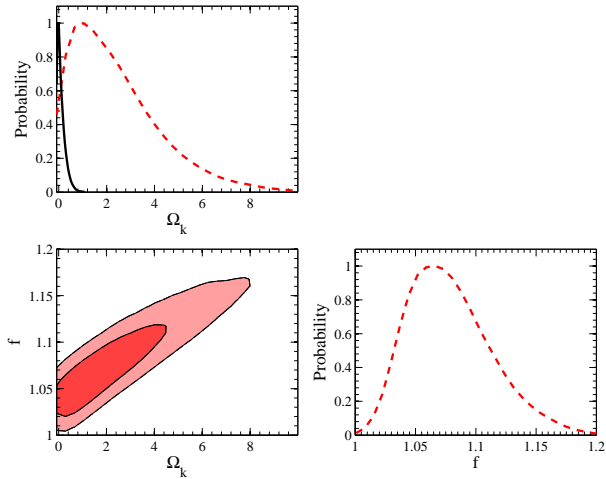}\\
    \caption{The 1-D and 2-D marginalized distributions with $1\sigma$ and $2\sigma$ contours for the parameters $\Omega_{\rm k}$ and $f$ constrained from the 23 SGL samples which were used for model Ia in \citet{Rasanen2015}. For comparison, we also show the one-dimensional distribution of $\Omega_{\rm k}$ by fixing $f=1$ (black solid line).}\label{fig:prl_omk}
\end{figure}

\subsection{SIE Model}

We start from the SIE model of the SGL system. In this model, we only need one free parameter $f$ to describe the SGL system using Equation (\ref{distanceRatio_obs}).

Firstly, we try to reproduce the numerical results of model Ia in \citet{Rasanen2015} which were obtained from a small sample of SGL system. Following their steps, we take $f=1$ and assign an error of 2\% on $\theta_{\rm E}$ and a minimum error of 5\% on $\sigma_{\rm c}$. In figure \ref{fig:prl_omk}, the black solid line in the one-dimensional distribution plot of $\Omega_{\rm k}$ is the result we obtain. The 95\% C.L. upper limit of the curvature: $\Omega_{\rm k}<0.84$, which is similar with the result in the previous work \citep{Rasanen2015}.

Then, we take the uncertainty of $f$ into account in the analysis. Different from the method in \citet{Rasanen2015}, in which they assigned an extra Gaussian error of 20\% on $d_{\rm ls}/d_{\rm s}$ from $f^2$, we treat the parameter $f$ as a free parameter and obtain the constraints on both $f$ and the curvature $\Omega_k$ from the SGL data simultaneously. The obtained one-dimensional and two-dimensional constraints on $\Omega_{\rm k}$ and $f$ are plotted in Figure \ref{fig:prl_omk}. We find that the constraint on the curvature is quite different from the result presented in \citet{Rasanen2015}. The 95\% upper limit of the curvature is:
\begin{equation}
  \Omega_{\rm k}<7.02~,
\end{equation}
which is much weaker than that obtained in the case with $f=1$. The reason is that the curvature is strongly correlated with the parameter $f$ of the SGL system, as shown in the left-bottom panel of Figure \ref{fig:prl_omk}. And the SGL data do not favor the standard SIS model at more than $2\sigma$ confidence level:
\begin{equation}\label{eq:prl_f}
  f=1.079\pm0.034~~(68\%~{\rm C.L.})~.
\end{equation}
If we force to set the parameter $f=1$, the obtained constraint on $\Omega_{\rm k}$ will be obviously biased. The constraining power of the SGL data is significantly overestimated. Furthermore, we also notice that our result is much weaker than that obtained for model Ib in \citet{Rasanen2015} which included an extra Gaussian error of 20\% on $d_{\rm ls}/d_{\rm s}$ from $f^2$. The reason is that including an extra Gaussian error from $f^2$ does not fully take the effect of $f$ into account. The strong correlation we find here is totally neglected in the analysis. Since this strong degeneracy between $f$ and $\Omega_{\rm k}$ is broken in their analysis, the constraint on $\Omega_{\rm k}$ becomes significantly tight. Furthermore, the current SGL data can already give very good constraint on $f$, see equation (\ref{eq:prl_f}), the uncertainty of $f^2$ is much smaller than 20\%.

\begin{figure}[t]
    \centering
    \includegraphics[width=0.5\textwidth]{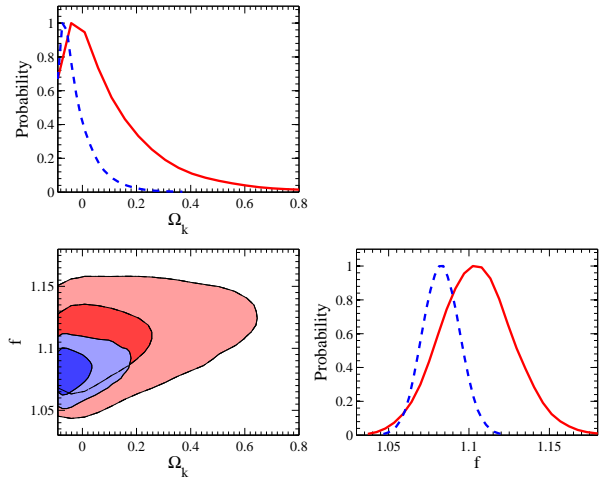}\\
    \caption{The 1-D and 2-D marginalized distributions with $1\sigma$ and $2\sigma$ contours for the parameters $\Omega_{\rm k}$ and $f$ constrained from the 83 SGL samples with (red) and without (blue) including the intrinsic scatter $\sigma_{\rm int}$ in the analysis.}\label{fig:cs_omk}
\end{figure}

Now we use the latest observation with 83 SGL samples to perform the global analysis. In Figure \ref{fig:cs_omk} we show the one-dimensional and two-dimensional constraints on $\Omega_{\rm k}$ and $f$ from the SGL data (blue dashed lines). Apparently, this new SGL data provide much stronger constraining power on $\Omega_{\rm k}$ and $f$ than the above small SGL sample used in \citet{Rasanen2015}, since the strong degeneracy between $\Omega_{\rm k}$ and $f$ is partly broken. Consequently, the constraints on $\Omega_{\rm k}$ and $f$ are much tighter than above:
\begin{eqnarray}
  \Omega_{\rm k}&<&0.16 ~~(95\% {\rm ~C.L.})~,\nonumber\\
  f&=&1.083\pm0.011 ~~(68\% {\rm ~C.L.})~.
\end{eqnarray}
We also check the minimal $\chi^2$ value of the best-fit model and find that $\chi^{2}_{\rm min}=\chi^2_{\rm SN} + \chi^2_{\rm SGL} = 545 + 243 = 788$. The $\chi^2$ of SN data looks normal, $\chi^2_{\rm SN}/{\rm d.o.f.}=545/580=0.94$, while the SGL data give too large $\chi^2$ value, $\chi^2_{\rm SGL}/{\rm d.o.f.}=243/83=2.93$. This might imply that there are some unknown uncertainties on the SGL samples.

In order to take this unknown effect into account in the analysis, we use the D'Agostini's likelihood \citep{DAgostini2005}:
\begin{align}\label{D_like}
\nonumber \mathcal{L}_D(\Omega_{\rm k},f,\sigma_{\rm int})\propto
&\prod_i\frac{1}{\sqrt{\sigma_{\rm int}^2+[\Delta(\sigma_{\rm c,i})]^2}}\\
&\times \exp\left[-\frac{[{\sigma_{\rm c,i}({\rm th})} - {\sigma_{\rm c,i}({\rm obs})}]^2}{2(\sigma_{\rm int}^2+[\Delta(\sigma_{\rm c,i})]^2)}\right]~,
\end{align}
where $\sigma_{\rm int}$ is the intrinsic scatter which represents any other unknown uncertainties except for the observational statistical ones, ${\sigma_{\rm c}({\rm th})}$ and ${\sigma_{\rm c}({\rm obs})}$ are the theoretical prediction and observation of the central stellar velocity dispersion, and $\Delta(\sigma_{\rm c})$ is the observed statistical error bar of SGL samples. By maximizing the D'Agostini's likelihood, or, equivalently, by minimizing the $\chi^2$, we could obtained the best-fit intrinsic scatter $\sigma_{\rm int}$ and its uncertainty. Then, we put a top-hat prior on $\sigma_{\rm int}$ and use the equation
\begin{equation}
  \chi^2=\sum_{\rm i=1}^{83}\frac{[{\sigma_{\rm c,i}({\rm th})} - {\sigma_{\rm c,i}({\rm obs})}]^2}{(\sigma_{\rm int}^2+[\Delta(\sigma_{\rm c,i})]^2)}~,
\end{equation}
to perform the whole calculations.

In Figure \ref{fig:cs_omk} we also show the one-dimensional and two-dimensional constraints on $\Omega_{\rm k}$ and $f$ from the SGL data with the intrinsic scatter $\sigma_{\rm int}$ included (red solid lines). Apparently, because the presence of the intrinsic scatter, the constraints on $\Omega_{\rm k}$ and $f$ become weaker, namely
\begin{eqnarray}\label{limit:f:all}
  \Omega_{\rm k}&<&0.60 ~~(95\% {\rm ~C.L.})~,\nonumber\\
  f&=&1.105\pm0.030 ~~(68\% {\rm ~C.L.})~,\nonumber\\
  \sigma_{\rm int}&=&31.8\pm4.2  ~~(68\% {\rm ~C.L.})~.
\end{eqnarray}
When comparing with the CMB and baryon acoustic oscillation measurements, the constraint on the curvature from the SGL data is very weak. However, this is a model-independent constraint based only on geometrical optics, thus could be the useful complementary to model specific analyses. On the other hand, similar with above analysis, the standard SIS model with $f=1$ is strongly disfavored by the SGL data at more than $3\sigma$ confidence level. More importantly, the minimal $\chi^2_{\rm SGL}$ now is about 85, and we obtain the reduced value $\chi^2_{\rm SGL}/{\rm d.o.f.}=1.03$ consequently. Therefore, in the following calculations, we always include the intrinsic scatter $\sigma_{\rm int}$ to represent any other unknown uncertainties of the SGL data.

\begin{figure}[t]
    \centering
    \includegraphics[width=0.5\textwidth]{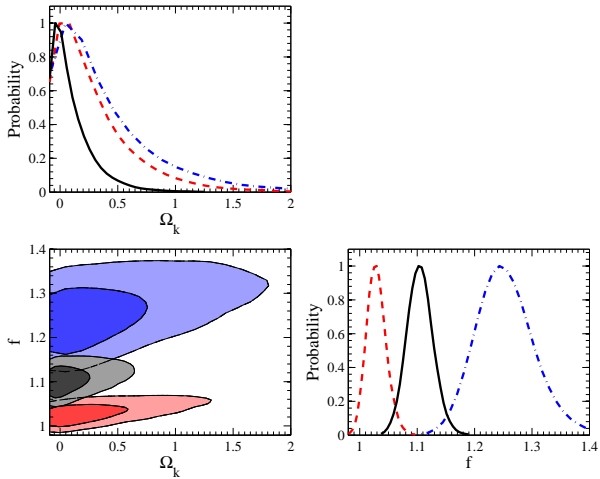}\\
    \caption{The 1-D and 2-D marginalized distributions with $1\sigma$ and $2\sigma$ contours for the parameters $\Omega_{\rm k}$ and $f$ constrained from the two subsamples with $\sigma_{\rm c}<240\,$km/s (blue dash-dotted lines) and $\sigma_{\rm c}>240\,$km/s (red dashed lines), respectively. For comparison, we also show the constraints from the whole 83 SGL samples (black solid lines).}\label{fig:cs_omk_sigma}
\end{figure}

In order to understand the interesting constraint on the parameter $f$ better, we first use the SGL systems in each survey to constrain $f$ separately. As we said before, in this SGL subsample, there are 20 samples from BELLS, 57 samples from SLACS and 6 samples from SL2S. Therefore, we use 20 samples from BELLS and 57 samples from SLACS to constrain $f$ and obtain the 68\% C.L. constraints, respectively:
\begin{eqnarray}
  f_{\rm BELLS}&=&2.06\pm0.37~,\nonumber\\
  f_{\rm SLACS}&=&1.07\pm0.03~,
\end{eqnarray}
which imply that the standard SIS model ($f=1$) is still ruled out at more than $2\sigma$ confidence level in each survey. Interestingly, we find that the median value of $f$ from BELLS is far away from the unity. However, due to the lack of SGL sample, the error bar of $f$ from the BELLS data is very large. In meanwhile, the constraint on $f$ from SLACS is more similar with that from all data [eq. (\ref{limit:f:all})]. The constraining power on $f$ mainly comes from the samples of SLACS survey.

Then, we divide this SGL data into two parts according the centric stellar velocity dispersion $\sigma_c$ of SGL system: $\sigma_{\rm c}<240\,$km/s ($n=41$ lenses) and $\sigma_{\rm c}>240\,$km/s ($n=42$ lenses). In Figure \ref{fig:cs_omk_sigma} we show the 1-D and 2-D marginalized distributions with $1\sigma$ and $2\sigma$ contours for the parameters $\Omega_{\rm k}$ and $f$ constrained from these two subsamples. Firstly, we can see clearly, due to the smaller SGL data in two subsamples, the upper limit constraints on the curvature are weaker: $\Omega_{\rm k}<1.69$ and $\Omega_{\rm k}<1.22$ at 95\% confidence level, from the small velocity dispersion subsample and large velocity dispersion subsample, respectively. Secondly, we find that the obtained constraints on the parameter $f$ are quite different from these two subsamples. The small velocity dispersion subsample favors a large value of $f$:
\begin{equation}
  f=1.25\pm0.05 ~~(68\% {\rm ~C.L.})~,
\end{equation}
which means the standard SIS model with $f=1$ is still ruled out by the data at high significance. By contrary, the large velocity dispersion subsample gives the different constraint on the parameter $f$:
\begin{equation}
  f=1.03\pm0.02 ~~(68\% {\rm ~C.L.})~.
\end{equation}
The standard SIS model is consistent with this large velocity dispersion subsample within 95\% confidence level. The larger value of velocity dispersion $\sigma_{\rm c}$ the subsample has, the more favored the standard SIS model with $f=1$ is.

Finally, we also separate this SGL data into two parts according the redshift of the lens galaxy: $z_{\rm l}<0.24$ ($n=39$ lenses) and $z_{\rm l}>0.24$ ($n=44$ lenses). Different from the above analysis, here we find that the constraints on $f$ are consistent with each other at about $1\sigma$ confidence level, namely:
\begin{eqnarray}
  f_{\rm z_{\rm l}<0.24}&=&1.10\pm0.04~~(68\% {\rm ~C.L.})~,\nonumber\\
  f_{\rm z_{\rm l}>0.24}&=&1.17\pm0.06~~(68\% {\rm ~C.L.})~.
\end{eqnarray}
The standard SIS model ($f=1$) is still ruled out at more than $2\sigma$ confidence level in each subsample.

\subsection{Extended SIE Model}

Besides the standard SIE model with one free parameter $f$, in our analysis we also consider the more complex SGL model. As we know, the measurement of central velocity dispersion $\sigma_{\rm c}$ can provide a model-dependent dynamical estimate of the mass, based on the assumption of the power-law mass density profile $\rho(r)$ and luminosity density of stars $\nu(r)$:
\begin{eqnarray}
  \rho(r) &=& \rho_0\left(\frac{r}{r_0}\right)^{-\alpha}~, \\
  \nu(r) &=& \nu_0\left(\frac{r}{r_0}\right)^{-\delta}~,
\end{eqnarray}
where $r$ is the spherical radial coordinate from the lens center. Therefore, besides the parameters $a_1$, $a_2$, $\Omega_{\rm k}$ and $\sigma_{\rm int}$, we have two more free parameters, $\alpha$ and $\delta$. Following \citet{Cao2016}, we can obtain the expression of the observed velocity dispersion:
\begin{eqnarray}\label{sigma_alpha_delta}
\nonumber \sigma_{\rm c}^2 &=& \left(\frac{c^2}{4}\frac{d_{\rm s}}{d_{\rm ls}}\theta_{\rm E}\right)\frac{2}{\sqrt{\pi}(\xi-2\beta)} \left( \frac{\theta_{\rm ap}}{\theta_{\rm E}}\right)^{2-\alpha}\\
&\times&\left[\frac{\lambda(\xi)-\beta\lambda(\xi+2)}{\lambda(\alpha)\lambda(\delta)}\right]
\frac{\Gamma{\left(\frac{3-\xi}{2}\right)}}{\Gamma{\left(\frac{3-\delta}{2}\right)}}~,
\end{eqnarray}
where $\beta$ is an anisotropy parameter to characterize the anisotropic distribution of three-dimensional velocity dispersion and has a Gaussian distribution $\beta=0.18\pm0.13$, based on the well-studied sample of nearby elliptical galaxies from \citet{gerhard2001}, $\theta_{\rm ap}$ is the spectrometer aperture radius, $\xi$ has the notation $\xi=\alpha+\delta-2$, and $\lambda(x)=\Gamma(\frac{x-1}{2})/\Gamma(\frac{x}{2})$ denotes the ratio of Euler's gamma functions. Finally, we have the $\chi^2$ equation in the calculations
\begin{equation}
  \chi^2=\sum_{\rm i=1}^{83}\left(\frac{\sigma_{\rm c,i}(z_{\rm l,i},z_{\rm s,i},\theta_{\rm E,i},\theta_{\rm ap,i};\alpha,\delta,\beta,\sigma_{\rm int}) - \sigma_{\rm c,i}({\rm obs})}{\Delta\sigma_{\rm c,i}}\right)^2~.
\end{equation}

\begin{figure}[t]
    \centering
    \includegraphics[width=0.5\textwidth]{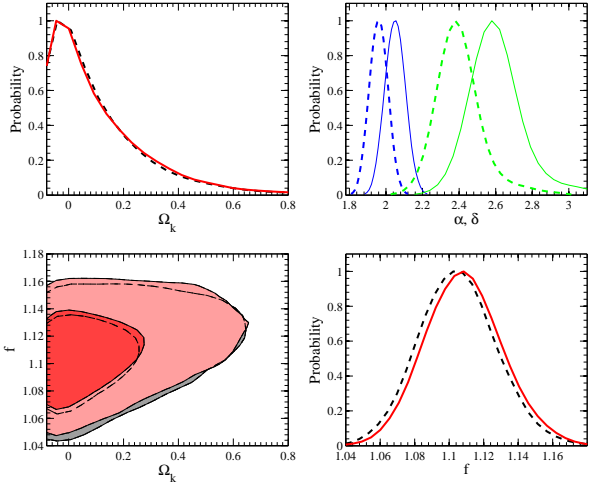}\\
    \caption{The 1-D and 2-D marginalized distributions with $1\sigma$ and $2\sigma$ contours for the parameters $\Omega_{\rm k}$ and $f$, as well as the power-law indexes $\alpha$ (blue) and $\delta$ (green) with (thin solid lines) and without (thick dashed lines) considering the seeing effect, constrained from the whole 83 SGL samples. For comparison, we also show the constraints on $\Omega_{\rm k}$ and $f$ by using the simple SIE model (black lines).}\label{fig:cs_omk_model}
\end{figure}

In Figure \ref{fig:cs_omk_model} we show the 1-D and 2-D marginalized distributions with $1\sigma$ and $2\sigma$ contours for the parameters $\Omega_{\rm k}$ and the power-law indexes $\alpha$ and $\delta$ constrained from the SGL data. The constraint on the curvature is identical with that obtained in the standard SIE model: $\Omega_{\rm k}<0.60$ at 95\% confidence level. The power indexes $\alpha$ and $\delta$ can also be well-constrained, namely the 68\% C.L. limits are:
\begin{eqnarray}
  \alpha &=& 1.97\pm0.04~, \\
  \delta &=& 2.40\pm0.13~,
\end{eqnarray}
in which the constraint of $\alpha$ is consistent with the SIS model at 95\% confidence level, while the $\delta$ constraint is ruled out the SIS model at more than $3\sigma$ confidence level. Note that, the slope $\delta$ can be directly measured from the observations, and the average mean value is $\delta=2.39$ with $1\sigma$ error bar $0.05$ \citep{Schwab10}, which is consistent with the result we obtain here. In the calculation, the minimal $\chi^2$ we obtain is about $84$ and $\chi^2_{\rm SGL}/{\rm d.o.f.}=1.03$, which is quite similar with those in the SIE model with the free parameter $f$.

We remark here that the above fitting results were obtained without considering the seeing effect. Considering the effects of aperture ($\theta_{ap}$) with atmospheric blurring ($\sigma_{\rm atm}$) and luminosity-weighted averaging (see \citet{Schwab10,Cao2016} for details), the constraints become: $\alpha=2.05\pm0.05$ and $\delta=2.61\pm0.15$ at $68\%$ confidence level, as shown in upper-right panel of figure \ref{fig:cs_omk_model}, which is consistent with previous works \citep{Koopmans2009,Sonnenfeld2013b,Oguri2014,Cao2016}.

Furthermore, we can also derive the parameter $f'$ by using
\begin{equation}\label{derived_f}
  f'=\left\{\frac{2}{\sqrt{\pi}(\xi-2\beta)}\left[\frac{\lambda(\xi)-\beta\lambda(\xi+2)}{\lambda(\alpha)\lambda(\delta)}\right]
\frac{\Gamma{\left(\frac{3-\xi}{2}\right)}}{\Gamma{\left(\frac{3-\delta}{2}\right)}}\right\}^{-1/2}~,
\end{equation}
and obtain the constraint from the SGL data:
\begin{equation}
  f'=1.108\pm0.030 ~~(68\% {\rm ~C.L.})~.
\end{equation}
Note that this term $f'$ is different from the parameter $f$ in the standard SIE model by a term $(\theta_{\rm ap}/\theta_{\rm E})^{2-\alpha}$. However, this neglected term is very close to the unity when $\alpha\simeq2$. Therefore, the posterior distribution of $f'$ in this model and one of $f$ in the SIE model are almost identical, see the right-bottom panel of Figure \ref{fig:cs_omk_model}.

If we assume that the radial profile of the luminosity density $\nu(r)$ follows that of the total mass density $\rho(r)$, namely $\alpha=\delta=\gamma$, we can obtain the constraint:
\begin{equation}
  \gamma = 2.04\pm0.02 ~~(68\% {\rm ~C.L.})~,
\end{equation}
which is consistent with the standard SIS model at $2\sigma$ level. And the 68\% C.L. limit on the derived term is $f'=1.00\pm0.01$, which is quite different from those results above. The reason is that when setting $\alpha=\delta=\gamma$, the term $f'$ in equation (\ref{derived_f}) is limited to be not far away from the unity. As we know, the whole SGL data do not favor the model with $f=1$. Therefore, when the term $f'$ is close to unity by force, the minimal $\chi^2_{\rm SGL}$ we obtain is much larger than those in the above models, $\chi^2_{\rm min,\gamma} - \chi^2_{\rm min,SIE} \simeq 53$. This assumption with $\alpha=\delta=\gamma$ has been ruled out by the current SGL data at very high significance.

\section{Discussion and Summary}\label{sec:discussion}

Since the growing importance of SGL system in research of astronomy, investigating the statistical properties of SGL systems becomes more and more important. In this paper, we use the latest SGL data observed by four surveys: SLACS, BELLS, LSD and SL2S to study the curvature of universe model-independently, based on the basic distance sum rule method, as well as the statistical properties of SGL systems in detail. Here we summarize our main conclusions in more detail:
\begin{itemize}
  \item Based on the standard SIS model with one free parameter $f$, we reproduce the analysis of \citet{Rasanen2015} on the constraint of the curvature from a small sample of SGL data. When we fix the parameter $f=1$, we obtain the similar constraint on the curvature with that in the previous work. However, when we set $f$ as a free parameter, due to the strong correlation between $\Omega_{\rm k}$ and $f$, the constraint on the curvature is significantly enlarged, namely the 95\% C.L. upper limit $\Omega_{\rm k}<7.02$. In the meanwhile, we also obtain the limit on the parameter $f$: $f=1.079\pm0.034$ (68\% C.L.), which implies that the standard SIS model with $f=1$ has been ruled out at more than $2\sigma$ confidence level. This result is different from that obtained in \citet{Rasanen2015}, because in their analysis they did not fully take the uncertainty of $f$ into account, but only assigned a Gaussian error of 20\% from $f^2$. This will bring the bias on the determination of $\Omega_{\rm k}$ and overestimate the constraining power of the SGL data on the curvature.
  \item We perform the global analysis on the curvature and the parameter $f$ by using the latest SGL data and find that this new SGL data can significantly improve the constraint on the curvature by a factor of 10, namely $\Omega_{\rm k}<0.60$ (95\% C.L.), when we introduce the intrinsic scatter, $\sigma_{\rm int}$, which represents any other unknown uncertainties, in the analysis. On the other hand, the constraint on the parameter $f$ is also improved: $f=1.105\pm0.030$ (68\% C.L.), which tells us that the SGL data do not favor the SIS model at more than $3\sigma$ confidence level. We also use the SGL samples in each survey to constrain $f$ separately and find that the most constraining power comes from the SLACS survey. Due to the lack of samples in BELLS survey, the limit on $f$ is very weak.
  \item In order to understand this result on $f$ better, we divide the whole SGL sample into two parts according the centric stellar velocity dispersion $\sigma_c$ of SGL system: $\sigma_{\rm c}<240\,$km/s and $\sigma_{\rm c}>240\,$km/s. Due to the smaller SGL data in two subsamples, the upper limit constraints on the curvature are weaker. More interestingly, the standard SIS model is consistent with the constraint of $f$ from the large velocity dispersion subsample within 95\% confidence level: $f=1.03\pm0.02$ ($2\sigma$ C.L.). The larger value of velocity dispersion $\sigma_{\rm c}$ the subsample has, the more favored the standard SIS model with $f=1$ is. We also divide the sample into two pars according the redshift of the lens galaxy: $z_{\rm l}<0.24$ and $z_{\rm l}>0.24$ and find that the constraints on $f$ are consistent with each other at about $1\sigma$ confidence level and different from the unity at more than $2\sigma$ confidence level.
  \item Besides the standard SIE model with one free parameter $f$, in our analysis we also consider the more complex SGL model by assuming the power-law mass density profile $\rho(r)=\rho_0(r/r_0)^{-\alpha}$ and luminosity density of stars $\nu(r)=\nu_0(r/r_0)^{-\delta}$. Using the whole SGL data, we obtain the constraints on the power-law indexes: $\alpha = 1.95\pm0.04$ and $\delta = 2.40\pm0.13$ at 68\% confidence level, which is consistent with the direct measurement from the observations on $\delta$. Comparing with the SIE model, we also obtain the constraint on the derived parameter $f'$, which is almost identical with the constraint on $f$ in the SIE model.
  \item We also assume that the radial profile of the luminosity density $\nu(r)$ follows that of the total mass density $\rho(r)$, namely $\alpha=\delta=\gamma$. This model strongly suggests $f'$ is very close to the unity which is disfavored by the SGL data. Therefore, the minimal $\chi^2_{\rm SGL}$ we obtain is much larger than those in the above models, $\chi^2_{\rm min,\gamma} - \chi^2_{\rm min,SIE} \simeq 53$. This assumption with $\alpha=\delta=\gamma$ has been ruled out by the current SGL data.
\end{itemize}

\section*{Acknowledgements}

J.-Q. Xia is supported by the National Youth Thousand Talents Program and the National Science Foundation of China under grant No. 11422323. G.-J. Wang, S.-X. Tian and Z.-H. Zhu are supported by the the NSFC under grant No. 11373014. H. Yu is supported by the National Basic Research Program of China (973 Program, grant No. 2014CB845800). Z.-X. Li is supported by the NSFC under grant No. 11505008. Shuo Cao is supported by the NSFC under grant No. 11503001. The research is also supported by the Strategic Priority Research Program ``The Emergence of Cosmological Structures'' of the Chinese Academy of Sciences, Grant No.XDB09000000, and the NSFC
under grant nos. 11633001 and 11690023.


\begin{thebibliography}{99}

\bibitem[Audren et al.(2013)]{Audren2013} Audren, B., Lesgourgues, J., Benabed, K., \& Prunet, S.\ 2013, JCAP, 2, 001

\bibitem[Audren(2014)]{Audren2014} Audren, B.\ 2014, MNRAS, 444, 827

\bibitem[Auger et al.(2009)]{Auger2009} Auger, M. W., et al.\ 2009, ApJ, 105, 1099

\bibitem[Bernstein(2006)]{Bernstein2006} Bernstein, G.\ 2006, ApJ, 637, 598

\bibitem[Bolton et al.(2008)]{Bolton2008} Bolton, A. S., et al.\ 2008, ApJ, 682, 964

\bibitem[Brownstein et al.(2012)]{Brownstein2012} Brownstein, et al.\ 2012, ApJ, 744, 41

\bibitem[Cai et al.(2016)]{Cai2016} Cai, R.-G., Guo, Z.-K., \& Yang, T.\ 2016, PRD, 93, 043517

\bibitem[Cao et al.(2012)]{Cao2012} Cao, S., Pan, Y., Biesiada, M., Godlowski, W., \& Zhu, Z.-H.\ 2012, JCAP, 3, 016

\bibitem[Cao et al.(2015)]{Cao2015} Cao, S., Biesiada, M., Gavazzi, R., Pi{\'o}rkowska, A., \& Zhu, Z.-H.\ 2015, ApJ, 806, 185

\bibitem[Cao et al.(2016)]{Cao2016} Cao, S., Biesiada, M., Yao, M., \& Zhu, Z.-H.\ 2016, MNRAS, 461, 2192

\bibitem[Chae et al.(2004)]{Chae2004} Chae, K.-H., Chen, G., Ratra, B., \& Lee, D.-W.\ 2004, ApJL, 607, L71

\bibitem[Clarkson et al.(2007)]{Clarkson2007} Clarkson, C., Cort{\^e}s, M., \& Bassett, B.\ 2007, JCAP, 8, 011

\bibitem[D'Agostini(2005)]{DAgostini2005} D'Agostini G.\ 2005, arXiv:physics/0511182

\bibitem[Efstathiou(2014)]{Efstathiou2014} Efstathiou, G.\ 2014, MNRAS, 440, 1138

\bibitem[Gerhard et al.(2001)]{gerhard2001} Gerhard, O., Kronawitter, A., Saglia, R. P., \& Bender, R.\ 2001, AJ, 121, 1936

\bibitem[Holanda et al.(2016)]{Holanda2016} Holanda, R.~F.~L., Busti, V.~C., \& Alcaniz, J.~S.\ 2016, JCAP, 2, 054

\bibitem[Humphreys et al.(2013)]{reid2013} Humphreys, E.M.L., Reid, M.J., Moran, J.M., Greenhill, L.J., \& Argon, A.L.\ 2013, ApJ, 775, 13


\bibitem[Kassiola \& Kovner(1993)]{Kassiola93} Kassiola, A., \& Kovner, I. 1993, ApJ, 417, 450
\bibitem[Keeton \& Kochanek(1998)]{Keeton98} Keeton, C. R., \& Kochanek, C. S. 1998, ApJ, 495, 157



\bibitem[Kochanek et al.(2000)]{Kochanek2000} Kochanek, C.~S., Falco, E.~E., Impey, C.~D., et al.\ 2000, ApJ, 543, 131

\bibitem[Koopmans \& Treu(2003)]{Koopmans2003} Koopmans, L.V.E. \& Treu, T.\ 2003, ApJ, 583, 606

\bibitem[Koopmans et al.(2009)]{Koopmans2009} Koopmans, L.V.E. et al.\ 2009, ApJL, 703, L51

\bibitem[Kormann et al.(1994)]{Kormann94} Kormann, R., Schneider, P., \& Bartelmann, M. 1994, A\&A, 284, 285

\bibitem[Lewis \& Bridle(2003)]{cosmomc} Lewis, A. \& Bridle, S.\ 2002, PRD, 66, 103511

\bibitem[Li et al.(2014)]{Li2014} Li, Y.-L., Li, S.-Y., Zhang, T.-J., \& Li, T.-P.\ 2014, ApJL, 789, L15

\bibitem[Li et al.(2016)]{Li2016} Li, X.-L., Cao, S., Zheng, X.-G., Li, S., \& Biesiada, M.\ 2016, Research in Astronomy and Astrophysics, 16, 015

\bibitem[Liang et al.(2008)]{Liang2008} Liang, N., Xiao, W.~K., Liu, Y., \& Zhang, S.~N.\ 2008, ApJ, 685, 354-360
\bibitem[Liao et al.(2016)]{Liao2016} Liao, K., Li, Z., Cao, S., et al.\ 2016, ApJ, 822, 74
\bibitem[Lin et al.(2016)]{Lin2016} Lin, H.-N., Li, X., \& Chang, Z.\ 2016, MNRAS, 455, 2131

\bibitem[Liu \& Wei(2015)]{Liu2015} Liu, J., \& Wei, H.\ 2015, General Relativity and Gravitation, 47, 141

\bibitem[Ofek et al.(2003)]{Ofek2003} Ofek, E.~O., Rix, H.-W., \& Maoz, D.\ 2003, MNRAS, 343, 639

\bibitem[Oguri et al.(2012)]{Oguri2012} Oguri, M. et al.\ 2012, AJ, 143, 120

\bibitem[Oguri et al.(2014)]{Oguri2014} Oguri, M., Rusu, C.E., \& Falco, E.E.\ 2014, MNRAS, 439, 2494

\bibitem[Peebles(1993)]{Peebles1993} Peebles, P.~J.~E.\ 1993, Principles of Physical Cosmology by P.J.E.~Peebles.~Princeton University Press, 1993.~ISBN: 978-0-691-01933-8, P336


\bibitem[Planck Collaboration et al.(2015)]{Planck2015} Planck Collaboration, Ade, P.~A.~R., Aghanim, N., et al.\ 2015, arXiv:1502.01589

\bibitem[R{\"a}s{\"a}nen et al.(2015)]{Rasanen2015} R{\"a}s{\"a}nen, S., Bolejko, K., \& Finoguenov, A.\ 2015, Physical Review Letters, 115, 101301

\bibitem[Ratnatunga et al.(1999)]{Ratnatunga1999} Ratnatunga, K.~U., Griffiths, R.~E., \& Ostrander, E.~J.\ 1999, AJ, 117, 2010

\bibitem[Riess et al.(2011)]{riess2011} Riess, A.G. et al.\ 2011, ApJ, 730, 119

\bibitem[Rusin et al.(2002)]{Rusin02} Rusin, D., et al. 2002, MNRAS, 330, 205

\bibitem[Saha(2000)]{Saha00} Saha, P. 2000, AJ, 120, 1654

\bibitem[Schwab et al.(2010)]{Schwab10} Schwab, J., Bolton, A. S., \& Rappaport, S. A. 2010, ApJ, 708, 750

\bibitem[Sonnenfeld et al.(2013a)]{Sonnenfeld2013a} Sonnenfeld, A., Gavazzi, R., Suyu, S. H., Treu, T., \& Marshall, P. J.\ 2013a, ApJ, 777, 97

\bibitem[Sonnenfeld et al.(2013b)]{Sonnenfeld2013b} Sonnenfeld, A., Treu, T., Gavazzi, R., Suyu, S. H., Marshall, P. J., Auger, M. W., \& Nipoti, C.\ 2013b, ApJ, 777, 98

\bibitem[Suzuki et al.(2012)]{Suzuki2012} Suzuki, N., Rubin, D., Lidman, C., et al.\ 2012, ApJ, 746, 85

\bibitem[Treu \& Koopmans(2002)]{Treu2002} Treu, T., \& Koopmans, L.V.E.\ 2002, ApJ, 575, 87

\bibitem[Treu \& Koopmans(2004)]{Treu2004} Treu, T., \& Koopmans, L.V.E.\ 2004, ApJ, 611, 739

\bibitem[Treu et al.(2006)]{Treu2006} Treu, T., Koopmans, L.~V., Bolton, A.~S., Burles, S., \& Moustakas, L.~A.\ 2006, ApJ, 640, 662

\bibitem[Vonlanthen et al.(2010)]{Vonlanthen2010} Vonlanthen, M., R{\"a}s{\"a}nen, S., \& Durrer, R.\ 2010, JCAP, 8, 023

\bibitem[Walsh et al.(1979)]{Walsh1979} Walsh, D., Carswell, R.~F., \& Weymann, R.~J.\ 1979, Nature, 279, 381

\bibitem[Wang et al.(2016)]{Wang2016} Wang, J.~S., Wang, F.~Y., Cheng, K.~S., \& Dai, Z.~G.\ 2016, AAP, 585, A68

\bibitem[Yu \& Wang(2016)]{Yu2016} Yu, H., \& Wang, F.~Y.\ 2016, arXiv:1605.02483

\bibitem[Zhu \& Wu(1997)]{Zhu1997} Zhu, Z.-H., \& Wu, X.-P.\ 1997, AAP, 326, L9


\bibitem[Zhu(2000)]{Zhu2000} Zhu, Z.-H.\ 2000, Modern Physics Letters A, 15, 1023

\end{thebibliography}
\end{document}